\documentclass[prd,aps,preprint,tightenlines,nofootinbib,superscriptaddress]{revtex4}
\usepackage{mathrsfs}
\usepackage{amsfonts}
\usepackage{amsmath}
\usepackage{array}
\usepackage{verbatim}
\usepackage{epsfig}
\usepackage{graphicx}

\begin{document}
\title{Soft Gluon Resummations in Dijet Azimuthal Angular Correlations at the Collider}

\author{Peng Sun}
\affiliation{Nuclear Science Division, Lawrence Berkeley National
Laboratory, Berkeley, CA 94720, USA}
\author{C.-P. Yuan}
\affiliation{Department of Physics and Astronomy, Michigan State University,
East Lansing, MI 48824, USA}
\author{Feng Yuan}
\affiliation{Nuclear Science Division, Lawrence Berkeley National
Laboratory, Berkeley, CA 94720, USA}

\begin{abstract}
We derive all order soft gluon resummation in dijet azimuthal angular correlation
in hadronic collisions at the next-to-leading logarithmic level. The relevant coefficients
for the Sudakov resummation factor, the soft and hard factors, are calculated.
The theory predictions agree well with the experimental data from D0 Collaboration
at the Tevatron. This provides a benchmark calculation for the transverse momentum
dependent QCD resummation for jet productions in hadron collisions and can be readily
applied at the CERN LHC.
\end{abstract}

\maketitle

{\it Introduction.} Jet production in high energy scattering is one of the basic
hard QCD processes in modern particle and nuclear physics~\cite{Sterman:1977wj}.
It has been used as a golden probe for perturbative QCD
studies and non-trivial hadron physics. In addition, jet physics
has played an important role in electro-weak physics and
searching for new physics beyond the standard
model at high energy colliders, such as the Tevatron and the
CERN Large Hadron Collider (LHC).
Meanwhile, formulating jet production in hadronic processes is also one
of the most challenging perturbative computations, and has
attracted strong theoretical interests in the last few
decades~\cite{Li:2011hy,Ridder:2013mf,deFlorian:2013qia}.

Among jet production processes, dijet production is one of
the special processes. This is not only because it is easy to
be identified experimentally, but also because it is an important channel to
search for new physics~\cite{Abazov:2009ac}.
In dijet events, the two jets are
produced mainly in the back-to-back configuration
in the transverse plane,
\begin{equation}
A+B\to Jet_1+Jet_2+X \ , \label{dijet}
\end{equation}
where $A$ and $B$ represent the two incoming hadrons
with momenta $P$ and $\overline{P}$, respectively,
the azimuthal angle between the two jets
is defined as $\phi=\phi_1-\phi_2$ with $\phi_{1,2}$
being the azimuthal angles of the two jets.
There have been comprehensive analyses of
the azimuthal angular correlation (or decorrelation)
in dijet events produced at
hadron colliders~\cite{Abazov:2004hm,Khachatryan:2011zj,daCosta:2011ni}.
In the leading order naive parton picture, the Born diagram
contributes to a Delta function at $\phi=\pi$.
One gluon radiation will lead to
a singular distribution around $\phi=\pi$,
which will persist at even higher orders.
This divergence arises when the
total transverse momentum of dijet (imbalance) is much smaller
than the individual jet momentum,
$q_\perp=|\vec{P}_{1\perp}+\vec{P}_{2\perp}|\ll |P_{1\perp}|\sim |P_{2\perp}|\sim P_J$,
where large logarithms appear in every order
of perturbative calculations. These large logs
are normally  referred as the Sudakov logarithms,
$\alpha_s^i\ln^{2i-1}(P_J^2/q_\perp^2)$.
Therefore, a QCD resummation has to be
included in order to have a reliable theoretical
prediction. Since it involves a transverse momentum
$q_\perp$, the resummation formalism adopted this process is
similar to that for low transverse momentum electroweak
boson (or Higgs boson) production:
the transverse momentum dependent (TMD) or
Collins-Soper-Sterman (CSS) resummation formalism~\cite{Collins:1984kg}.
However, due to the fact that the colored
final state of dijet will induce additional soft gluon radiations,
the resummation of dijet production is much more
complicated.
In the literature, the leading double logarithmic
contribution for dijet azimuthal correlation has been
derived~\cite{Banfi:2008qs,Mueller:2013wwa}, where it was found that each incoming parton
contributes half of its color charge to the leading double
logarithmic Sudakov resummation factor. In this paper,
we will go beyond the double logarithmic approximation to
perform the resummation calculation at the next-to-leading logarithm (NLL)
level. More importantly, we will construct a theoretical framework
in the TMD (or CSS) resummation formalism to describe
jet production processes which will have great impact on the LHC physics.
The NLL resummation for this observable has been considered in 
Ref.~\cite{Banfi:2008qs} from different perspective. We also
note that a Mont Carlo event generator has been developed
to study this observable~\cite{Hautmann:2008vd}, where
a $k_t$-dependent parton shower was applied.

The methodology and technique of our calculations
follow the original studies on the threshold
resummation of colored final state particle production in hadronic collisions
by Sterman {\it et al.}~\cite{Kidonakis:1997gm},
where it was found that a matrix form of the resummation
formula has to be applied~\cite{Botts:1989kf}.
We find that the same conclusion also holds for studying dijet production
using the TMD resumamtion formlism.
Recent studies for the TMD resummation for heavy
quark pair production in  hadronic collisions also found
similar matrix form~\cite{Zhu:2012ts,zhu13}.

Our resummation formula can be summarized as
\begin{eqnarray}
\frac{d^4\sigma}
{dy_1 dy_2 d P_J^2
d^2q_{\perp}}=\sum_{ab}\sigma_0\left[\int\frac{d^2\vec{b}_\perp}{(2\pi)^2}
e^{-i\vec{q}_\perp\cdot
\vec{b}_\perp}W_{ab\to cd}(x_1,x_2,b_\perp)+Y_{ab\to cd}\right] \ ,
\end{eqnarray}
where the first term $W$ contains all order resummation
and the second term $Y$ comes from the fixed order corrections;
$\sigma_0$ represents normalization of the
differential cross section, $y_1$ and $y_2$ are rapidities of the two jets, $P_J$ the
jet transverse momentum, and $q_\perp$ the imbalance transverse momentum between
the two jets as defined above. All order resummation for $W$ from each
partonic channel $ab\to cd$ can be written as
\begin{eqnarray}
W_{ab\to cd}\left(x_1,x_2,b\right)&=&x_1\,f_a(x_1,\mu=b_0/b_\perp)
x_2\, f_b(x_2,\mu=b_0/b_\perp) e^{-S_{\rm Sud}(Q^2,b_\perp)} \nonumber\\
&\times& \textmd{Tr}\left[\mathbf{H}_{ab\to cd}
\mathrm{exp}[-\int_{b_0/b_\perp}^{Q}\frac{d
\mu}{\mu}\mathbf{\gamma}_{}^{s\dag}]\mathbf{S}_{ab\to cd}
\mathrm{exp}[-\int_{b_0/b_\perp}^{Q}\frac{d
\mu}{\mu}\mathbf{\gamma}_{}^{s}]\right]\ ,\label{resum}
\end{eqnarray}
where $Q^2=\hat s=x_1x_2S$, which represents the hard momentum scale, $b_0=2e^{-\gamma_E}$,
$f_{a,b}(x,\mu)$ are parton distributions for the incoming
partons $a$ and $b$, $x_{1,2}=P_J\left(e^{\pm y_1}+e^{\pm y_2}\right)/\sqrt{S}$
are momentum fractions of the incoming hadrons carried by the partons.
In the above equation, the hard and soft factors $\mathbf{H}$ and $\mathbf{S}$
are expressed as matrices in the color space of partonic channel $ab\to cd$, and $\gamma^s$
are the associated anomalous dimensions for the soft factor (defined below). The Sudakov
form factor ${\cal S}_{Sud}$ resums the leading double logarithms and
the universal sub-leading logarithms,
\begin{eqnarray}
S_{\rm Sud}(Q^2,b_\perp,C_1,C_2)=\int^{Q^2}_{b_0^2/b_\perp^2}\frac{d\mu^2}{\mu^2}
\left[\ln\left(\frac{Q^2}{\mu^2}\right)A+B+D_1\ln\frac{Q^2}{P_J^2R_1^2}+
D_2\ln\frac{Q^2}{P_J^2R_2^2}\right]\ , \label{su}
\end{eqnarray}
where $R_{1,2}$ represent the cone sizes for the two jets.
Here the parameters $A$, $B$, $D_1$, $D_2$ can be expanded
perturbatively in $\alpha_s$. At one-loop order,
$A=C_A \frac{\alpha_s}{\pi}$,
$B=-2C_A\beta_0\frac{\alpha_s}{\pi}$ for gluon-gluon initial state,
$A=C_F \frac{\alpha_s}{\pi}$,
$B=\frac{-3C_F}{2}\frac{\alpha_s}{\pi}$ for quark-quark initial state,
and $A=\frac{(C_F+C_A) }{2}\frac{\alpha_s}{\pi}$,
$B=(\frac{-3C_F}{4}-C_A\beta_0)\frac{\alpha_s}{\pi}$ for gluon-quark initial state.
These coefficients $A$, at the one-loop order, agree with those found in the leading double
logarithmic analysis~\cite{Banfi:2008qs,Mueller:2013wwa}. In our numeric calculations, we will also
include $A^{(2)}$ contributions since they are associated with the incoming
parton distributions and are the same (after dividing by a factor of two)
as those for vector boson and Higgs
particle productions~\cite{deFlorian:2000pr}.

At the next-to-leading logarithmic level, the jet cone size enters as well.
That is the reason we have two additional factors in Eq.~(\ref{su}):
$D=C_A\frac{\alpha_s}{2\pi}$ for gluon jet and $D=C_F\frac{\alpha_s}{2\pi}$ for quark jet.
The cone size $R$ is introduced to regulate the collinear
gluon radiation associated with the final state jets. Only the soft
gluon radiation outside the jet cone contributes to the imbalance $q_\perp$ between the two jets,
and yields the logarithmic contribution in the form of
$\ln\left(\frac{Q^2}{P_J^2 R^2}\right)$.
This conclusion is independent of
the jet algorithm~\cite{Mukherjee:2012uz}.
It is interesting to note that the similar cone size dependence in Eq.~(\ref{su}) has also
been found in the threshold resummation for jet production in hadronic collisions~\cite{deFlorian:2013qia,Kidonakis:1997gm}.

In the following, we will explain briefly how we derive the above resummation
results and present the numeric comparison with experimental data.
The detailed derivations will be left for a separate publication.
We will also examine $\alpha_s$ expansion of the resummed
result against the full one-gluon radiation perturbative
calculations in the limit of $q_\perp\ll P_J$, i.e., close to the back-to-back
configuration $\phi=\pi$. This serves as an important cross check of our
resummation formula.

{\it TMD Distributions.}
The total transverse momentum of the dijet $q_\perp$ is perpendicular to
the beam direction of incoming hadrons. The most important
contributions come from the collinear and soft gluon radiations in the limit
of $q_\perp\ll P_J$.
The collinear gluon radiations associated with the incoming partons
are factorized into the relevant parton distributions.
In this paper, we assume this factorization is valid for dijet production. 
By comparing to the experimental data, we will be able to test the factorization
and investigate the factorization breaking effects in this process,
which has been extensively discussed in the literature in the last few
years~\cite{Collins:2007nk,Mulders:2011zt,Vogelsang:2007jk,Catani:2011st, Mitov:2012gt}.
The factorization breaking effects found in these studies emerge at order
$\alpha_s^3$. Hence, we only include up to $\alpha_s^2$ contribution
in the the resummation coefficient $A$ of Eq.~(\ref{su}).

To evaluate the gluon radiation contribution associated with the parton
distributions, we introduce the TMD parton distributions, following the
Ji-Ma-Yuan scheme. For example, for the gluon distribution
from hadron $A$, we have~\cite{Ji:2004wu},
\begin{eqnarray}
xg(x,k_\perp)&=&\int\frac{d\xi^-d^2\xi_\perp}{P^+(2\pi)^3}
    e^{-ixP^+\xi^-+i\vec{k}_\perp\cdot \vec\xi_\perp}\nonumber\\
    &&\times
    \left\langle P|{F_a^+}_\mu(\xi^-,\xi_\perp)
{\cal L}^\dagger_{vab}(\xi^-,\xi_\perp) {\cal L}_{vbc}(0,0_\perp)
F_c^{\mu+}(0)|P \right\rangle\ ,\label{gpdfu}
\end{eqnarray}
where $F^{\mu\nu}$ is the gauge field strength tensor, and
${\cal L}_v(\xi)=P\exp\left(-ig\int_{0}^{-\infty}
    d\lambda {v}\cdot A(\lambda { v} + \xi) \right)$ is the
gauge link in the adjoint representation, $A^\mu=-if_{abc}  A_c^\mu$.
The off-light-cone vector $v$ is introduced to regulate the
light-cone singularity associated with the TMD
distributions, $\zeta^2=(2v\cdot P)^2/v^2$. An evolution
equation can be derived for the TMD distributions respect to
$\zeta$,
\begin{equation}
\frac{\partial}{\partial\ln\zeta} xg(x,b_\perp,\zeta)=\left(K(\mu,b_\perp)+G(\zeta,\mu)\right)\times xg(x,b_\perp,\zeta) \ ,
\end{equation}
where $K$ and $G$ are the evolution kernel.
Similarly, we will introduce the TMD parton
distribution from incoming hadron $B$, which includes
another light-cone singularity regulator $\bar \zeta^2=(2\bar v\cdot \bar P)^2/\bar v^2$.
After resummation, the dependence on $v$ and $\bar v$ will
cancel out between TMD distributions and the soft factors, which we
will introduce in the following.

{\it Soft Factor.} The collinear gluon radiations
parallel to the final state two jets are factorized into the jet functions,
and do not contribute to the low transverse momentum imbalance $q_\perp$.
However, the soft gluon radiations will contribute. In this section,
we will evaluate the soft factor contribution.

For the color neutral particle production in hadronic collisions,
the soft factor can be easily constructed. For colored final state,
factorizing out the soft gluon radiation
is much more complicated, although the basic idea is
the same, i.e., applying the Eikonal approximation. So, for each incoming
and outgoing partons, the soft gluon radiation is factorized
into an associated gauge link moving along the parton momentum
direction.
Because of color entanglement among
the four partons in the $2\to 2$ process, gauge invariant
construction of the soft factor depends on the overall color
configurations. In our calculations, we follow the procedure
of Ref.~\cite{Kidonakis:1997gm}, where the soft gluon radiations are evaluated
on the orthogonal color basis. For example, in the
partonic process $gg\to q\bar q$, there are three color bases,
\begin{gather}
 C_1  = \delta^{a_1a_2} \delta_{a_3a_4} \, , \qquad  C_2
   = if^{a_1a_2c} \, T^c_{a_3a_4} \, , \qquad  C_3  =
  d^{a_1a_2c} \, T^c_{a_3a_4} \, .
\label{eq:colorstructures}
\end{gather}
where $a_{3,4} $ are the color indices for the finial state quark and antiquark, and
$a_{1,2} $ for the initial state gluons. For $qg\to qg$ channel, similar color bases
will be chosen; whereas there are 8 independent color structures for $gg\to gg$ channel.
The soft factor is constructed
as matrix elements on the above bases, and
\begin{eqnarray}
S_{IJ}&=&\int_0^\pi
\frac{d\phi_0}{\pi}\; C^{bb'}_{Iii'} C^{aa'}_{Jll'}\langle 0|{\cal L}_{
vcb'}^\dagger(b_\perp) {\cal
L}_{ vbc} (b_\perp){\cal L}_{\bar
vca'}^\dagger(0) {\cal
L}_{\bar vac}(0) {\cal L}_{n ji}^\dagger(b_\perp) {\cal
L}_{\bar n i'k}(b_\perp) {\cal L}_{\bar nkl}^\dagger (0) {\cal
L}_{nl'j} (0)  |0\rangle \ ,\label{soft}
\end{eqnarray}
where we have integrated out the azimuthal angle $\phi_1$ of the
leading jet and traded the relative azimuthal angle $\phi$ for $q_\perp$. In the above equation,
$I,J$ represent the color basis index,
$n$ and $\bar n$ represent final state quark and antiquark momentum
directions (for the above mentioned $gg\to q\bar q$ channel),
and $v$ and $\bar v$ for the initial state two momentum directions.
The soft factor is expressed by a $3\times 3$ matrix. Accordingly, the hard
factor can also be calculated on the same color bases and
expressed as a $3\times 3$ matrix.

Following the TMD formalism, we need to choose the
off-light-cone gauge links for the two incoming partons in Eq.~(\ref{soft}).
As mentioned in the Introduction, not all the soft gluon radiations
can contribute to the imbalance $q_\perp$ between the two jets. Those gluon
radiation inside the jet cone will be part of the jet and will not contribute
to the soft factor. In order to exclude these contributions, we can impose
a kinematic constraint on the phase space integral for the radiated gluon.
Equivalently, we find that it is much simpler to require an off-shellness of
$n$ and $\bar n$ in Eq.~(\ref{soft}) so that
$n^2=P_J^2R_1^2/Q^2$ and $\bar n^2=P_J^2R_2^2/Q^2$,
where $R_{1,2}$ are the cone sizes for the two
jets~\cite{deFlorian:2013qia,Kidonakis:1997gm}. By doing so,
we obtain the exact same leading logarithms of $R_{1,2}$ for the
soft factor as compared to the kinematic constraint calculation
in the small cone approximation.

The soft factor satisfies the
renormalization group equation~\cite{Kidonakis:1997gm}:
 \begin{eqnarray}
\frac{d}{d\ln\mu}S_{IJ}(\mu) &=&
-\Gamma_{IJ'}^{s\dag}S_{J'J}(\mu)-S_{IJ'}(\mu)\Gamma_{J'J}^{s}\ ,\label{eq:evolution}
\end{eqnarray}
where $\Gamma$ is the relevant anomalous dimension.
By solving the renormalization group
equation, we resum the large logarithms associated
with the soft factor. More specifically, we include the cone size dependent
term in Eq.~(\ref{su}) and keep the rest in the $\gamma^s$ term in
Eq.~(\ref{resum}).

Factorization implies that the differential
cross section contributions from the partonic processes
can be written as
\begin{eqnarray}
W_{ab\to cd}\left(x_i,b\right)=x_1\,f_a(x_1,b,\zeta^2,\mu^2,\rho)x_2 f_b(x_2,b,\bar\zeta^2,\mu^2,\rho)
 \textmd{Tr}\left[\mathbf{H}_{ab\to cd}(Q^2,\mu^2,\rho)\mathbf{S}_{ab\to cd}(b,\mu^2,\rho)\right] \nonumber \ ,
\end{eqnarray}
where the dependence on $\rho=(2v\cdot \bar v)^2/v^2\bar v^2$ and the factorization scale $\mu$ cancel
out among different factors.
To derive the final resummation results, we have to solve the
evolution equation for the parton distributions and the renormalization
group equation for the soft factor.
In particular, we will choose the factorization scale $\mu=Q$,
and evolve the parton distribution and soft factor from the
scale $1/b$ to $Q$. After this, we arrive at the resummation
results, as shown in Eqs.~(2) and (3). The detailed expressions for $\mathbf{H}$,
$\mathbf{S}$ and $\gamma^s$ will be given in a separate publication.
Two important cross checks are found in
the final results: (1) $\rho$ dependence cancels out;
(2) cone size $R$ dependence
is diagonal in the color basis matrix and can be removed from
the soft factor, which leads to a universal resummation
factor as shown in Eqs.~(3) and (4).

{\it Compare to the Experimental Data.}
Before we apply our resummation formula to compare with experimental data,
we would like to demonstrate the consistency between the perturbative expansion of
our resummation result and the fixed order calculation around $\phi=\pi$ .
In Fig.~1, we plot the normalized distribution of the dijet azimuthal angle $\phi$,
in which the largest transverse momentum ($p_T^{\rm max}$) of the jets falling
into the range of $100 < P_T^{\rm max} < 130\,{\rm GeV}$.
The solid curve is from a full leading order (LO) calculation (with three partons in the
final state) and the dashed curve is from the expansion of the $W_{ab\to cd}$ term
to the same order in strong coupling constant $\alpha_s$, labelled as
``singular'' piece.
In the numeric calculations, we used the CT10 next-to-leading order (NLO) parton
distribution functions~\cite{Gao:2013xoa}, and the jet cone size $R=0.7$~\cite{Abazov:2004hm}.
From this figure, we clearly see that the leading order expansion
of the resummation results agrees well with the fixed
order calculation in the limit $\phi\to \pi$. This demonstrates that
the resummation result captures the most important contribution from the fixed order
calculation.
Away from $\phi=\pi$ region, there will be difference, which
is included in the resummation calculation via the $Y$ term of Eq.~(2).

\begin{figure}[tbp]
\centering
\includegraphics[width=8cm]{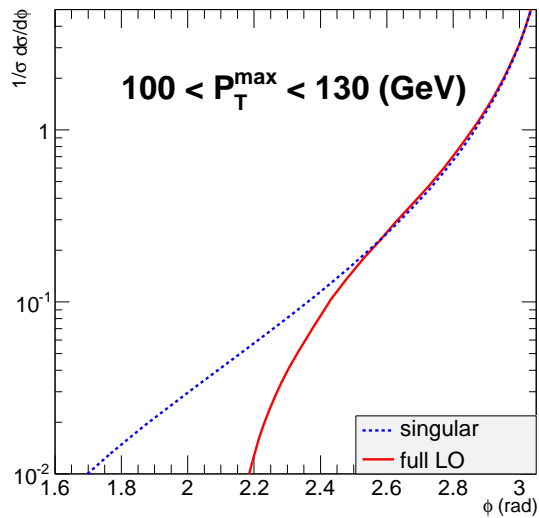}
\caption{Comparison between the $\alpha_s$ expansion of the
resummation calculation (dashed curve) and the full leading order (LO)
calculation (solid curve) in the region around $\phi = \pi$ for
$100 < P_T^{\rm max} < 130\,{\rm GeV}$, with the
same kinematics specified by the D0 Collaboration~\cite{Abazov:2004hm}.}
\label{asymptotic}
\end{figure}

In Fig.~2, we compare our resummation results
to the experimental data from D0 Collaboration at the Tevatron~\cite{Abazov:2004hm}.
We have included $A^{(1,2)}$, $B^{(1)}$ and $D_{1,2}^{(1)}$
in the Sudakov form factor. It is worthwhile to mention that
including $A^{(2)}$ improves the agreements with the
data, which may support the factorization we argue in this
paper. $B$ and $D$ coefficients are generally process-dependent.
Here, we only include their one-loop results.
It is desirable to obtain higher order
results and study their effects, which is however
beyond the scope of this paper.

\begin{figure}[tbp]
\centering
\includegraphics[width=8cm]{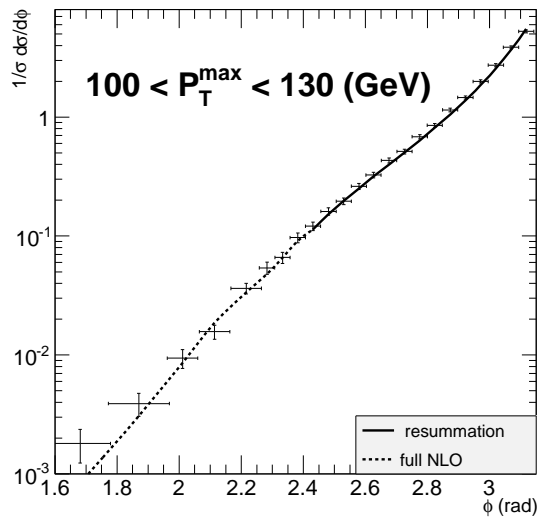}
\caption{Resummation results on dijet azimuthal correlations
at the Tevatron, compared to the experimental data from D0 Collaboration~\cite{Abazov:2004hm}.
Around $\phi=2.4$, the resummation result (solid curve)
is matched to a full NLO calculation (dashed curve)~\cite{Nagy:2001fj}.}
\label{exp}
\end{figure}

When Fourier transforming the $b_\perp$-expression to obtain the
transverse momentum distribution, we follow the $b_*$ prescription
of CSS resummation~\cite{Collins:1984kg}, {\it i.e.},
replacing $b_\perp$ by $b_*=b/\sqrt{1+b^2/b_{max}^2}$ in the calculation.
By doing so, we will also introduce the non-perturbative
form factors for the quarks and gluons from the initial states. In our calculations,
we have used $b_{max}=0.5\,{\rm GeV}^{-1}$, and the non-perturbative form factors
follow the parameterizations in Refs.~\cite{Sun:2012vc}. However, we would like
to emphasize that because the jet energy is so large that our final
results are not sensitive to the non-perturbative form factors at all.

From Fig.~\ref{exp}, we see that our resummation
results agree well with the experimental data, from $\phi$ near to $\pi$ down to
much smaller values. For smaller value of $\phi$
(away from the back-to-back configuration), the resummation calculations match to
the fixed order results at NLO~\cite{Nagy:2001fj}, which has
also been separately shown in Fig.~\ref{exp}.
We note that a full NLO calculation cannot describe
experimental data for $\phi\sim \pi$~\cite{Abazov:2004hm},
where the fixed order calculation becomes divergent.
Our resummation calculation, after being matched with
the NLO result (at $\phi$ around $2.4$ in this example),
clearly improves the theory prediction and can describe the experimental data
in a wider kinematic region.
This demonstrates
the importance of all order resummation in perturbative calculations
for these type of hard QCD processes.

{\it Conclusions.} In this paper, we have performed the QCD resummation
calculation for dijet azimuthal angular correlation in hadron collisions,
at the next-to-leading logarithmic level. The relevant resummation
coefficients are computed
at one-loop order.
We show that the perturbative expansion of our resummation
calculation agree well with the fixed order calculation
in the back-to-back correlation limit ($\phi$ around $\pi$).
After being matched with the full NLO calculation at a smaller
$\phi$ value, our resummation results can describe a much wider range
of the experimental data, particulary for $\phi$ around $\pi$ where
event rate dominates.
The agreements between the resummation results and the experimental data
encourage further developments along this direction. We will present the detailed
derivations and more phenomenological studies
in a future publication. We will also discuss jet algorithm dependence
of our results and the issues concerning the ``non-global" logarithms arising
from $\alpha_s^2$ calculations~\cite{Banfi:2008qs}.

Extension to other jets production processes will be interesting to carry out,
in particular, for those that involve Higgs boson in the final state, such as
Higgs plus one jet or two jets production
at the LHC. Applications to the dijet production in
heavy ion collisions and proton-nucleus collisions~\cite{Dominguez:2010xd}
are interesting to follow as well. We also notice that recently, the ATLAS collaboration at the LHC
has measured the dijet azimuthal correlation with large rapidity separation between the jets,
which has been interpreted as the BFKL resummation effects~\cite{Ducloue:2013bva}.
It will be worthwhile to compare these theoretical calculations with our results to pin
down the underlying physics mechanism for the observed phenomena.

We thank Al Mueller, Jianwei Qiu, Werner Vogelsang, and Bowen Xiao for interesting
discussions. We also thank Andrea Banfi and Mrinal Dasgupta for communications
concerning their results in Ref.~\cite{Banfi:2008qs}.
This work was partially supported by the U.S. Department of Energy
via grant DE-AC02-05CH11231, and by the U.S. National
 Science Foundation under Grant No. PHY-0855561.


\begin{thebibliography}{99}


\bibitem{Sterman:1977wj}
  G.~F.~Sterman and S.~Weinberg,
  Phys.\ Rev.\ Lett.\  {\bf 39}, 1436 (1977).

\bibitem{Li:2011hy}
  H.~-n.~Li, Z.~Li and C.~-P.~Yuan,
  Phys.\ Rev.\ Lett.\  {\bf 107}, 152001 (2011).

\bibitem{Ridder:2013mf}
  A.~Gehrmann-De Ridder, T.~Gehrmann, E.~W.~N.~Glover and J.~Pires,
  Phys.\ Rev.\ Lett.\  {\bf 110}, 162003 (2013).

\bibitem{deFlorian:2013qia}
  D.~de Florian, P.~Hinderer, A.~Mukherjee, F.~Ringer and W.~Vogelsang,
  Phys.\ Rev.\ Lett.\  {\bf 112}, 082001 (2014).

\bibitem{Abazov:2009ac}
see, for example,  V.~M.~Abazov {\it et al.}  [D0 Collaboration],
  Phys.\ Rev.\ Lett.\  {\bf 103}, 191803 (2009).

\bibitem{Abazov:2004hm}
  V.~M.~Abazov {\it et al.}  [D0 Collaboration],
  Phys.\ Rev.\ Lett.\  {\bf 94}, 221801 (2005).

\bibitem{Khachatryan:2011zj}
  V.~Khachatryan {\it et al.}  [CMS Collaboration],
  Phys.\ Rev.\ Lett.\  {\bf 106}, 122003 (2011).

\bibitem{daCosta:2011ni}
  G.~Aad {\it et al.}  [ATLAS Collaboration],
  Phys.\ Rev.\ Lett.\  {\bf 106}, 172002 (2011).

\bibitem{Collins:1984kg}
  J.~C.~Collins, D.~E.~Soper and G.~F.~Sterman,
  Nucl.\ Phys.\ B {\bf 250}, 199 (1985).

\bibitem{Banfi:2008qs}
  A.~Banfi, M.~Dasgupta and Y.~Delenda,
  Phys.\ Lett.\ B {\bf 665}, 86 (2008).


\bibitem{Mueller:2013wwa}
  A.~H.~Mueller, B.~-W.~Xiao and F.~Yuan,
  Phys.\ Rev.\ D {\bf 88}, 114010 (2013).

\bibitem{Hautmann:2008vd}
  F.~Hautmann and H.~Jung,
  JHEP {\bf 0810}, 113 (2008).

\bibitem{Kidonakis:1997gm}
  N.~Kidonakis and G.~F.~Sterman,
  Nucl.\ Phys.\ B {\bf 505}, 321 (1997);
  N.~Kidonakis, G.~Oderda and G.~F.~Sterman,
  Nucl.\ Phys.\ B {\bf 525}, 299 (1998).

\bibitem{Botts:1989kf}
  J.~Botts and G.~F.~Sterman,
  Nucl.\ Phys.\ B {\bf 325}, 62 (1989).

\bibitem{Zhu:2012ts}
  H.~X.~Zhu, C.~S.~Li, H.~T.~Li, D.~Y.~Shao and L.~L.~Yang,
  Phys.\ Rev.\ Lett.\  {\bf 110}, 082001 (2013);
   Phys.\  Rev.\  D {\bf 88}, 074004 (2013).

  \bibitem{zhu13}
  R.~Zhu, C.~Qiao, P.~Sun, F.~Yuan, to be published.

\bibitem{deFlorian:2000pr}
  D.~de Florian and M.~Grazzini,
  Phys.\ Rev.\ Lett.\  {\bf 85}, 4678 (2000).

\bibitem{Mukherjee:2012uz}
  A.~Mukherjee and W.~Vogelsang,
  Phys.\ Rev.\ D {\bf 86}, 094009 (2012).


\bibitem{Collins:2007nk}
  J.~Collins and J.~-W.~Qiu,
  Phys.\ Rev.\ D {\bf 75}, 114014 (2007).
  \bibitem{Mulders:2011zt}
  T.~C.~Rogers and P.~J.~Mulders,
  Phys.\ Rev.\ D {\bf 81}, 094006 (2010).
\bibitem{Vogelsang:2007jk}
  W.~Vogelsang and F.~Yuan,
  Phys.\ Rev.\ D {\bf 76}, 094013 (2007).
\bibitem{Catani:2011st}
  S.~Catani, D.~de Florian and G.~Rodrigo,
  JHEP {\bf 1207}, 026 (2012).

\bibitem{Mitov:2012gt}
  A.~Mitov and G.~Sterman,
  Phys.\ Rev.\ D {\bf 86}, 114038 (2012).

\bibitem{Ji:2004wu}
  X.~Ji, J.~P.~Ma and F.~Yuan,
  Phys.\ Rev.\ D {\bf 71}, 034005 (2005);
  JHEP {\bf 0507}, 020 (2005).

\bibitem{Nagy:2001fj}
  Z.~Nagy,
  Phys.\ Rev.\ Lett.\  {\bf 88}, 122003 (2002);
  Phys.\ Rev.\ D {\bf 68}, 094002 (2003).

\bibitem{Gao:2013xoa}
  J.~Gao, M.~Guzzi, J.~Huston, H.~-L.~Lai, Z.~Li, P.~Nadolsky, J.~Pumplin and D.~Stump {\it et al.},
  Phys.\ Rev.\ D {\bf 89}, 033009 (2014).

\bibitem{Sun:2012vc}
  F.~Landry, R.~Brock, P.~M.~Nadolsky and C.~P.~Yuan,
  Phys.\ Rev.\ D {\bf 67}, 073016 (2003);
  Phys.\ Rev.\ D {\bf 63}, 013004 (2001);
  P.~Sun, C.~-P.~Yuan and F.~Yuan,
  Phys.\ Rev.\ D {\bf 88}, 054008 (2013).

  \bibitem{Dominguez:2010xd}
  F.~Dominguez, B.~-W.~Xiao and F.~Yuan,
  Phys.\ Rev.\ Lett.\  {\bf 106}, 022301 (2011);
  F.~Dominguez, C.~Marquet, B.~-W.~Xiao and F.~Yuan,
  Phys.\ Rev.\ D {\bf 83}, 105005 (2011).


\bibitem{Ducloue:2013bva}
  B.~DuclouŽ, L.~Szymanowski and S.~Wallon,
  Phys.\ Rev.\ Lett.\  {\bf 112}, 082003 (2014).



\end{thebibliography}
\end{document}